\input{aipcheck}

\documentclass[
    ,final            
  ]
  {aipproc}

\layoutstyle{6x9}

\begin{document}

\title[The GeV to TeV view of SNR IC443: predictions for Fermi]{The GeV to TeV view of SNR IC443: predictions for Fermi}

\classification{98.38.Mz,95.55.Ka}
\keywords      {SNR (individual IC 443), $\gamma$-rays: observations, $\gamma$-rays: theory}

\author{Ana Y. Rodriguez Marrero}{
  address={arodrig@ieec.uab.es, Institut de Ci\`encies de l'Espai (IEEC-CSIC),
    Campus UAB,  Torre C5, \mbox{2a planta},
    08193 Barcelona, Spain}
}

\author{Diego F. Torres}{
  address={dtorres@ieec.uab.es}
}

\author{Elsa de Cea Del Pozo}{
  address={decea@ieec.uab.es}
}

\begin{abstract}
We present a theoretical model that explains the high energy 
phenomenology of the neighborhood of SNR IC 443, as observed with the Major Atmospheric Gamma Imaging Cherenkov (MAGIC) telescope and the Energetic Gamma-Ray Experiment Telescope (EGRET). 
We also discuss how the model can be tested with observations by the Fermi Gamma-ray Large Area Space Telescope.
We interpret  MAGIC J0616+225 as delayed TeV emission of cosmic-rays diffusing from IC 443 and interacting with a known cloud located at a distance of about 20 pc in the foreground of the remnant. 
This scenario naturally explains the displacement between EGRET and MAGIC sources, their fluxes, and their spectra. 
Finally, we predict how this context can be observed by Fermi.
 
\end{abstract}

\maketitle


\section{Introduction}

Recently,  MAGIC presented the results of observations towards SNR IC 443,
yielding to the detection of a new source of 
$\gamma$-rays, J0616+225 (Albert et al. 2007). This source is located at (RA,DEC)=(06$^\mathrm{h}$16$^\mathrm{m}$43$^\mathrm{s}$,
+22$^\circ$31'48''), with a  statistical positional error of 1.5', and 
a systematic error 
of 1'.
A simple power law was fitted to the measured
spectral points:
$
{ \mathrm{d}N_{\gamma}}/ ({\mathrm{d}A \mathrm{d}t \mathrm{d}E})
= (1.0 \pm 0.2) \times  10^{-11} 
\left({E}/{\mathrm{0.4 TeV}}\right)^{-3.1 \pm 0.3}
 \mathrm{cm}^{-2}\mathrm{s}^{-1}
\mathrm{TeV}^{-1},$
with quoted errors being statistical.  
No variability was found along the observation time
(over one year). No significant tails nor extended structure was found  at the MAGIC angular resolution.
 
MAGIC~J0616+225 is displaced with respect to the position of the non-variable (Torres et al. 2001) EGRET source 3EG J0617+2238 (Hartman et al. 1999). Indeed, the EGRET  central position is located directly towards the SNR, whereas the MAGIC source is 
south of it, close to the 95\% CL contour of the EGRET detection. As Albert et al. (2007) showed, the MAGIC source is located at the position of a giant cloud in front of the SNR, it would not be surprising if they are related, which we explore here.
The EGRET flux is (51.4$\pm$3.5) $\times 10^{-8}$ ph cm$^{-2}$ s$^{-1}$, with a photon spectral index of 2.01$\pm$0.06. 
Extrapolating the spectrum of the EGRET source into the VHE regime, we would obtain a higher flux and a harder spectrum than which was observed for MAGIC~J0616+225, supporting the view that a direct extrapolation of this and other EGRET measurements into the VHE range is not valid (Funk et al. 2008).

Here we present a theoretical model (see Aharonian \& Atoyan 1996, also Gabici \& Aharonian 2007) explaining the high energy phenomenology of  IC 443, making focus in the displacement between EGRET and MAGIC sources. Our interpretation of MAGIC  J0616+225 is that it is delayed TeV emission of cosmic-rays (CRs) diffusing from the SNR. Finally, we discuss how the model can be tested with observations with the Fermi Gamma-ray Large Area Telescope. 

\section{The SNR IC 443 in context of the model}

IC 443 is one of the most studied SNR region at all frequencies. It is an asymmetric shell-type SNR with a diameter of $\sim$45 arc
minutes (e.g., Fesen \& Kirshner 1980). Two half shells appear in optical and radio images (e.g, Braun \& Strom 1986, Leahy 2004, Lasker et al. 1990). 
The  interaction region, with evidence for multiple dense clumps,
is also seen in 2MASS images (e.g. Rho et al. 2001).
In radio, IC 443 has a spectral index of 0.36, and a flux density of 160~Jy at 
1~GHz (Green 2004).
Claussen et al.  (1997)
reported the presence of maser emission at 1720~MHz 
at $(l,b)\sim(-171.0,\, 2.9)$. Recently, Hewitt et al. (2006)
confirmed Claussen's et al. measurements and discovered 
weaker maser sources in the region of interaction.
%
IC 443 is a prominent X-ray source, observed with Rosat (Asaoka \& Aschenbach 1994), ASCA (Keohane 1997), XMM (Bocchino \& Bykov 2000, 2001, 2003, Bykov et al. 2005, Troja et al. 2006), and Chandra (Olbert et al. 2001, Gaensler et al. 2006). The works by Troja et al (2006) and
Bykov et al. (2008) summarize these observations.
In what follows we present some additional features of IC 443, relevant for our model.

{\it  Age:} IC 443 is agreed to have a middle-age of about $3 \times 10^4$ yrs. This age has been initially advocated by Lozinkaya (1981) and was later consistently obtained as a result of the SNR evolution model (Chevalier 1999).

{\it Distance:} Kinematical distances from optical systemic velocities span from 0.7 to 1.5 kpc (e.g., Lozinskaya 1981). The assumption that the SNR is associated with a nearby HII region, S249, implies a distance of $\sim 1.5-2.0$ kpc. Several authors claimed that the photometric distance is more reliable (e.g, Rosado et al. 2007), and concurrently with all other works on IC 443, we adopt its distance as 1.5 kpc (thus, 1 arcmin corresponds  to 0.44 pc).

{\it Energy of the explosion:} There is no clear indicator for $E_{51}$, the energy of the explosion in units of $10^{51}$ erg.
 Lacking a strong reason for other numerical assumption, we will assume that $E_{51}=1$, although to be conservative, we will subsequently assume that only 5\% of this energy is converted into relativistic 
CRs. Reasonable differences in our assumed value of $E_{51}$ are not expected to have any impact on this model.

{\it The molecular environment:}
Cornett et al. (1977) and DeNoyer (1979) were among the first to present detailed observations of molecular lines towards IC 443. Subsequently, Dickman et al. (1992), Seta et al. (1998), Butt et al. (2003) and Torres et al. (2003) among others, presented further analysis. These works conform the current picture for the environment of IC 443: a total mass of $\sim 1.1 \times 10^4$ M$_\odot$ mainly located in a quiescent 
cloud in front of the remnant (with linear scales of a few parsecs and densities of a few hundred particles cm$^{-3}$) that is absorbing optical and X-ray radiation (e.g., Lasker 1990, Troja et al. 2006), a scenario already put forward by Cornett et al. (1977).  
Dickman et al. (1992) estimated that 500-2000 M$_\odot$ are directly perturbed by the shock in the northern region of interaction, near the SNR itself.
Huang et al. (1986) found several clumps of molecular material along this interacting shell, with subparsec linear scales. Rosado et al. (2007) found inhomogeneities down to 0.007 pc. As it is usual, we will neglect these latter inhomogeneities when considering the propagation of CRs in the ISM, i.e. we thus assume an homogeneous medium of typical ISM density where CRs diffuse. Then, the molecular mass scenario is  a main giant cloud in front of the SNR containing most of the quiescent molecular material found in the region, and smaller cloud(s) totalizing the remaining mass located closer to the SNR.

\section{Diffusion of CRs from IC 443}

The spectrum of $\gamma$-rays generated through $\pi^0$-decay at
a source of proton density $n_{p}$ is 
$
   F_{\gamma}(E_{\gamma})=2\int^{\infty}_{E_{\pi}^{\rm min}}
({F_{\pi}(E_{\pi})}/{\sqrt{E_{\pi}^{2}-m_{\pi}^{2}}})
   \;dE_{\pi},
$
where 
$
E_{\pi}^{\rm
min}(E_{\gamma})=E_{\gamma}+ {m_{\pi}^{2}}/{4E_{\gamma}},
$
and
$
F_{\pi}(E_{\pi})=4\pi n_{p}\int^{E^{\rm max}_{p}}_{E^{\rm
min}_{p}} J_p(E) ({d\sigma_{\pi}(E_{\pi},\;E_{p})}/{dE_{\pi}})
\;dE_{p}.
$
Here, $d\sigma_{\pi}(E_{\pi},\;E_{p})/dE_{\pi}$ is the
differential cross-section for the production of $\pi^0$-mesons of
energy $E_{\pi}$ by a proton of energy $E_{p}$ in a $pp$
collision. For an study of different parameterizations of this cross section see Domingo-Santamaria \& Torres (2005) and Kelner et al. (2006). The limits of integration in the last expression are obtained by kinematic considerations (see e.g., Torres 2004). 
In these expressions we have implicitly neglected any possible gradient of cosmic-ray density in the cloud as well as in the cloud's gas number density.

The CR spectrum is  given by
$
   J_p(E,\;r,\;t)= [{c} \beta / {4\pi}] f,
$
where $f(E,\;r,\;t)$ is the distribution function of protons at
an instant $t$ and distance $r$ from the source. The distribution
function satisfies the radial-temporal-energy dependent diffusion equation (Ginzburg \&
Syrovatskii 1964):
$
   ({\partial f}/{\partial t})=({D(E)}/{r^2}) ({\partial}/{\partial
   r}) r^2 ({\partial f}/{\partial r}) + ({\partial}/{\partial
   E}) \, (Pf)+Q,
$
where $P=-dE/dt$ is the energy loss rate of the
particles, $Q=Q(E,\;r,\;t)$ is the source function, and $D(E)$
is the diffusion coefficient, for which we assume here that it depends only on the particle's energy. The energy loss rate are due to ionization and nuclear interactions, with the latter dominating over the former for energies larger than 1 GeV. The nuclear loss rate is $P_{\rm nuc} = E/\tau_{pp}$, with $\tau_{pp}=(n_p\, c \, \kappa \, \sigma_{pp} ) ^{-1}$ being the timescale for the corresponding nuclear loss, $\kappa \sim 0.45$ being the inelasticity of the interaction, and $\sigma_{pp}$ being the cross section (Gaisser 1990). 
Aharonian \& Atoyan (1996) presented a solution for the diffusion equation for an arbitrary energy loss term, diffusion coefficient, and impulsive  injection spectrum $f_{\rm inj}(E)$, such that  $Q(E,r,t) = N_0 f_{\rm inj}(E) \delta{\bar r} \delta(t)$. For the particular case in which $D(E)\propto E^\delta$ and $f_{\rm inj}\propto
E^{-\alpha}$, 
the general solution is 
$
  f(E, r,t) \sim ({N_0 E^{-\alpha}}/{\pi^{3/2} R_{\rm dif}^3}) \exp \left[ { - {(\alpha-1)t}/{\tau_{pp} }- ({R}/{R_{\rm dif}})^2} \right],
$
where $R_{\rm dif} = 2 ( D(E) t [\exp(t \delta / \tau_{pp})-1]/[t \delta / \tau_{pp}])^{1/2}$ stands for the radius of the sphere up to which the particles of energy $E$ have time to propagate after their injection.
In case of continuous injection of accelerated particles, given by $Q(E, \;
t)=Q_0 E^{-\alpha} {\cal T}(t)$, the previous solution needs to be convolved with the function ${\cal T}(t-t')$ in the time interval $0 \leq t' \leq t$. 
If the source is described by a Heavside function, 
 ${\cal T}(t)=\Theta (t)$ 
Atoyan et al. (1995) have found a general solution for the diffusion equation 
with arbitrary injection spectrum, which with the
listed assumptions and for times $t$ less than the energy loss time, leads to:
$
f(E,\;r,\;t)=({Q_0 E^{-\alpha}}/{4\pi D(E) r}) (
{2}/{\sqrt{\pi}})\int^{\infty}_{r/R_{\rm diff}} e^{-x^2}
dx.
$ 
We will assume that $\alpha=2.2$ and make use of these solutions in what follows.


Fig. 1 shows the current CR spectrum 
generated by IC 443 at two different distances from the accelerator, 10 (solid) and 30 (dashed) pc. The SNR is 
considered both as a continuous accelerator with
a relativistic proton power of  $L_p = 5 \times 10^{37}$ erg s$^{-1}$ 
(the proton luminosity is such that the energy injected into relativistic CRs through the SNR age is $5\times 10^{49}$ erg), and an impulsive injector with the same total power (injection of high energy particles occur in a much shorter time than the SNR age). 
The horizontal line in Fig. 1 marks the CR spectrum near Earth, so that the excess of CRs in the SNR environment can be seen. For this example, the diffusion coefficient at 10 GeV, $D_{10}$, was chosen as 10$^{26}$ cm$^2$ s$^{-1}$, with $\delta=0.5$. CRs propagate through the ISM, assumed to have a typical density. In the scale of Fig. 1, curves for $n_{ISM}=0.5, 1, 5,$ and 10 cm$^{-3}$ would be superimposed, so that $n_{ISM}$ becomes an irrelevant parameter in this range (this stems from the fact that the timescale for nuclear loss $\tau_{pp}$ obtained with the densities considered for the interestellar medium, $n_{ISM}$, is orders of magnitude larger than the age of the accelerator). Differences between the different kind of accelerators assumed are also minimal for the SNR parameters.

Fig. 2 shows the result for the $\gamma$-ray emission coming from the cloud located at the position of the MAGIC source, when we assume it lies at different distances in front of IC 443. The giant cloud mass is assumed (consistently with observations) as 8000 M$_\odot$. The accelerator properties and power of IC 443 are as in Fig. 1, in each case. Fluxes are given for an ISM propagation in a medium of $n=1$ cm$^{-3}$, although again we have checked this is not a relevant parameter as discussed above. We find that clouds located from $\sim$20 to $\sim$30 pc produce an acceptable match  to MAGIC data. In the case of a more impulsive accelerator, the VHE predicted spectra is slightly steeper than that produced in the continuous case at the same distance, so that it provides a correspondingly better fit to the MAGIC spectrum.
Fig. 2 also shows, apart from MAGIC data, EGRET measurements of the neighborhood of IC 443. We recall that these two sources are not located at the same place, what we emphasize using different symbols.  Fig. 2 shows that there is plenty of room for a cloud the size of that detected in front of IC 443 to generate the MAGIC source and not a co-spatial EGRET detection. In the case of Fermi, measurement of this region will allow us to constrain the separation between the SNR and the cloud, since for some distances a Fermi detection is also predicted. The existence of a VHE source without counterpart at lower energies is the result of 
diffusion of the high-energy CRs from the SNR shock, which is an energy dependent process leading to an increasing deficit of low energy protons the farther is the distance from the accelerator.

To clarify our previous assertion, and since our solution to the diffusion-loss equation is a function of time, we show the evolution of the flux along the age of the SNR. 
In Fig. 3 we show the integrated photon flux coming from the position of the giant cloud as a function of time above 100 MeV and 100 GeV in the impulsive case. Different qualities of the accelerator (impulsive or continuous) produce a rather comparable picture. At the age of the SNR (the time at which we observe) Fermi should see a source only for the closest separations. On the contrary, the integrated photon fluxes above 100 GeV present minimal deviations, and a MAGIC source is always expected.

\begin{figure}
\centering
\includegraphics[width=0.8\columnwidth,trim=0 5 0 10]{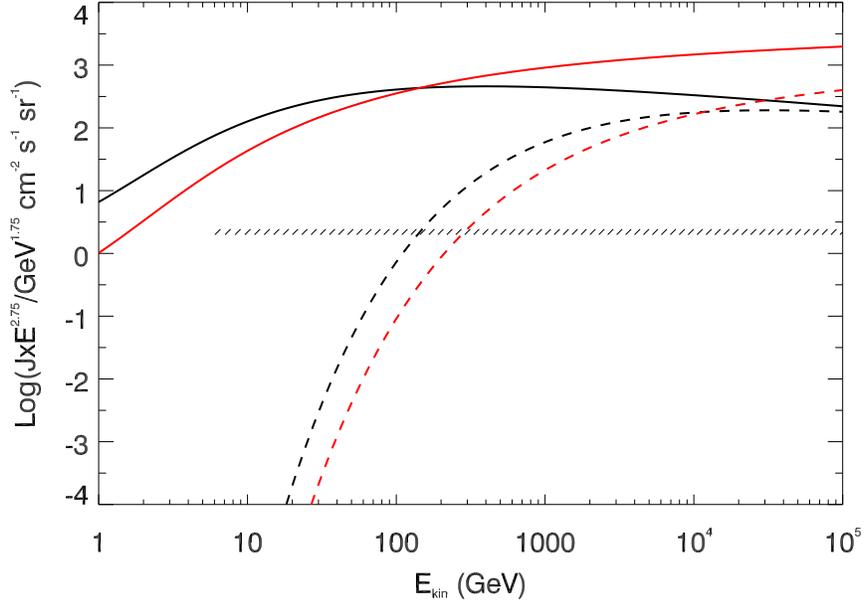}
\caption{Current CR spectrum 
generated by IC 443
at two different distances, 10 (solid) and 30 (dashed) pc, at the age of the SNR.
Two types of accelerator are considered, one providing a continuous injection (black)
and other providing a more impulsive injection of CRs (red). 
The horizontal line marks the CR spectrum near the Earth. The y-axis units have been chosen to emphasize the excess of CRs in the SNR environment.}
\end{figure}

\begin{ltxfigure}
\centering
\includegraphics[width=0.8\columnwidth,trim=0 5 0 10]{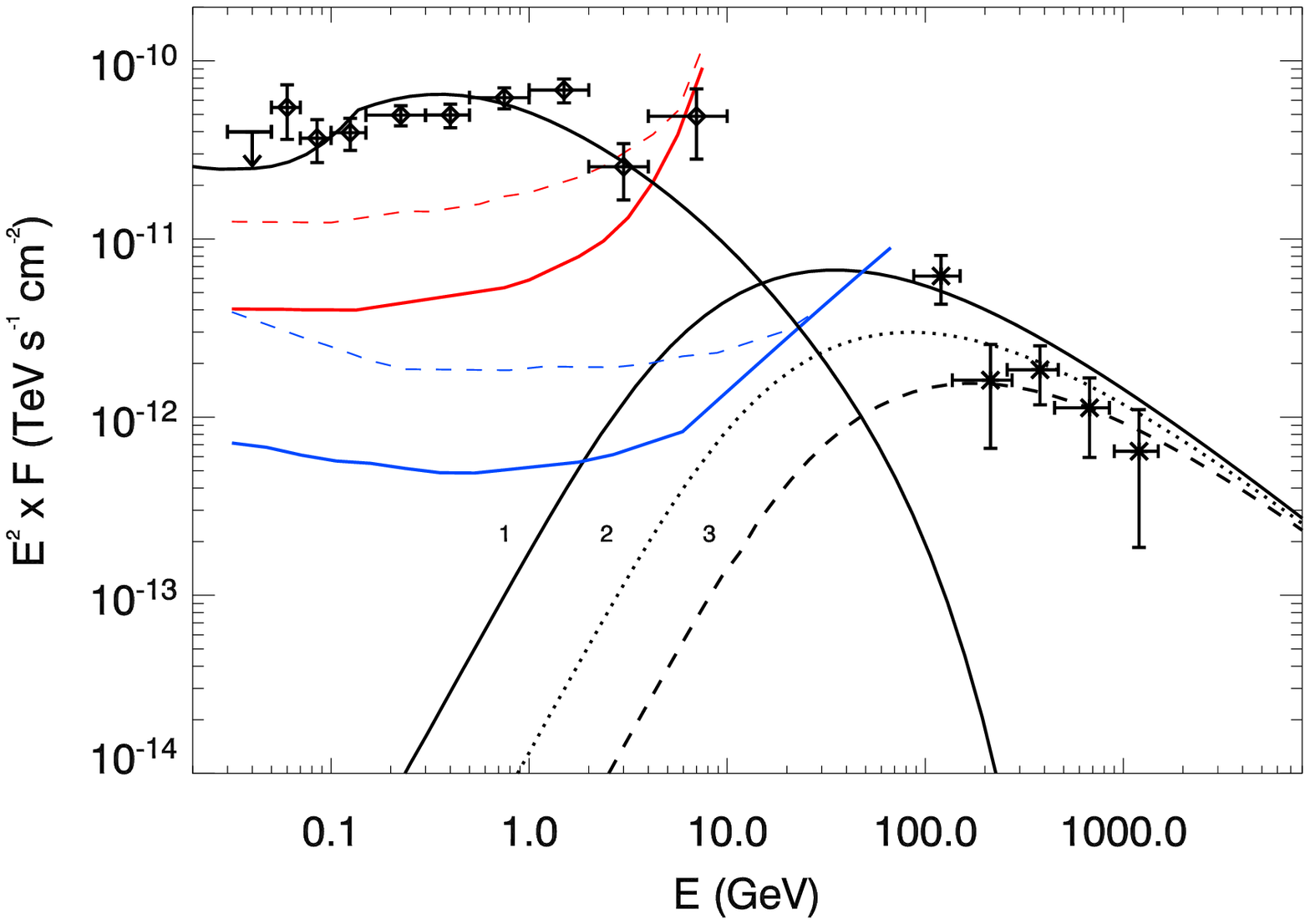}
\includegraphics[width=0.8\columnwidth,trim=0 5 0 10]{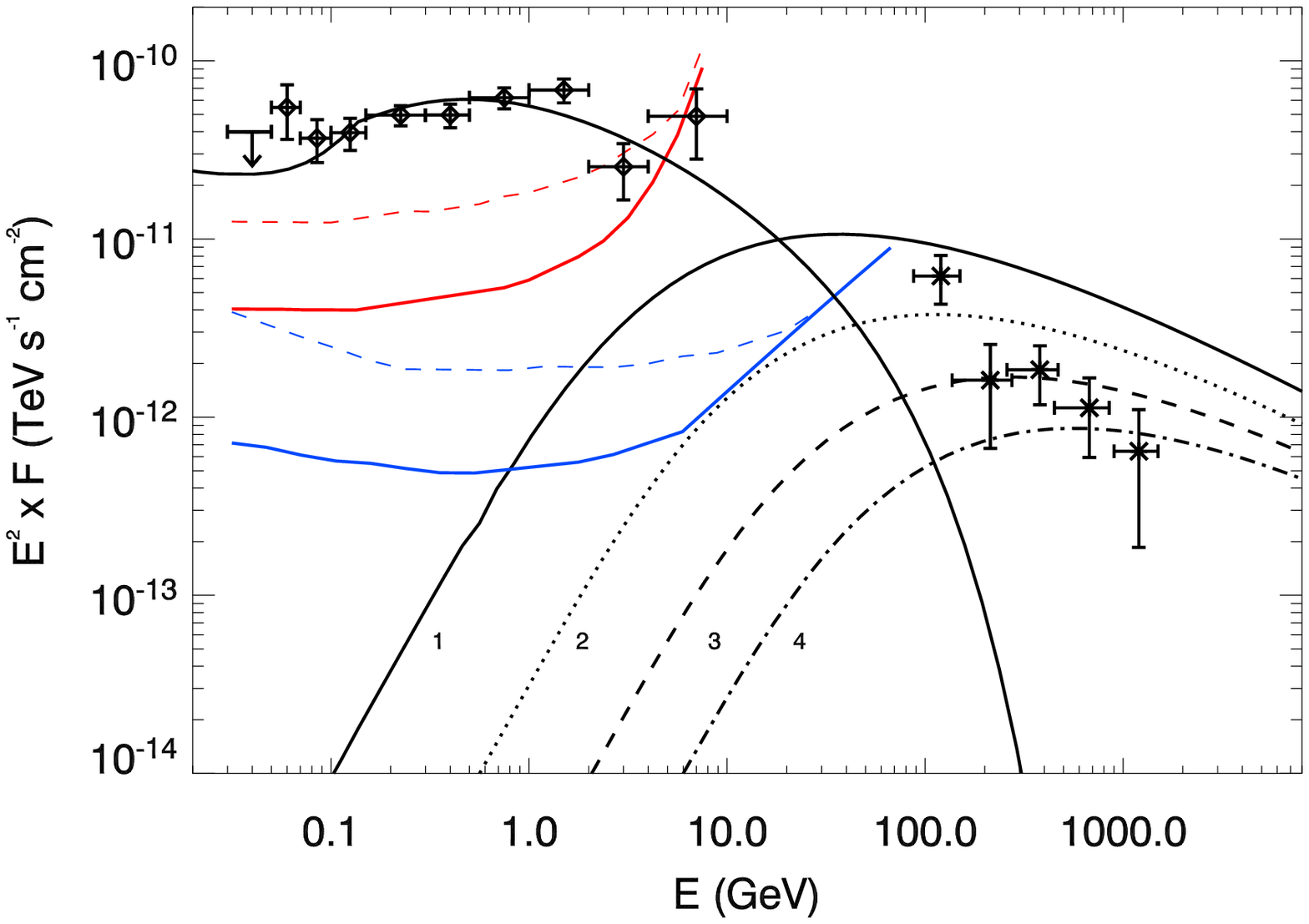}
\caption{MAGIC and EGRET measurement of the neighborhood of IC 443 (stars and squares, respectively) as compared with model predictions. 
The top (bottom) panel shows the results for an impulsive (continuous) case.
At the MAGIC energy range, the top panel curves show the predictions for a cloud of 8000 M$_\odot$ located at 20 (1), 25 (2), and 30 (3) pc, whereas they correspond to 
15 (1), 20 (2), 25 (3), and 30 (4) pc in the bottom panel. At the EGRET energy range, the curve shows the prediction for a few hundred M$_\odot$ located at 3--4 pc.
The EGRET sensitivity curves (in red) are shown for the whole lifetime of the mission for the Galactic anti-centre (solid), which received the largest exposure time and has a lower level of diffuse $\gamma$-ray emission, and 
for a typical position in the Inner Galaxy (dashed), more
dominated by diffuse $\gamma$-ray background. 
The Fermi sensitivity curves (in blue) (taken 
from http://www-glast.slac.stanford.edu/software/IS/glast\_latperformance.html)
show the  1-year sky-survey sensitivity for the Galactic North pole, again a position with low diffuse 
emission (solid), and for a typical position in the Inner Galaxy (dashed).
}
\end{ltxfigure}

\begin{figure}
\centering
\includegraphics[width=0.8\columnwidth,trim=0 5 0 10]{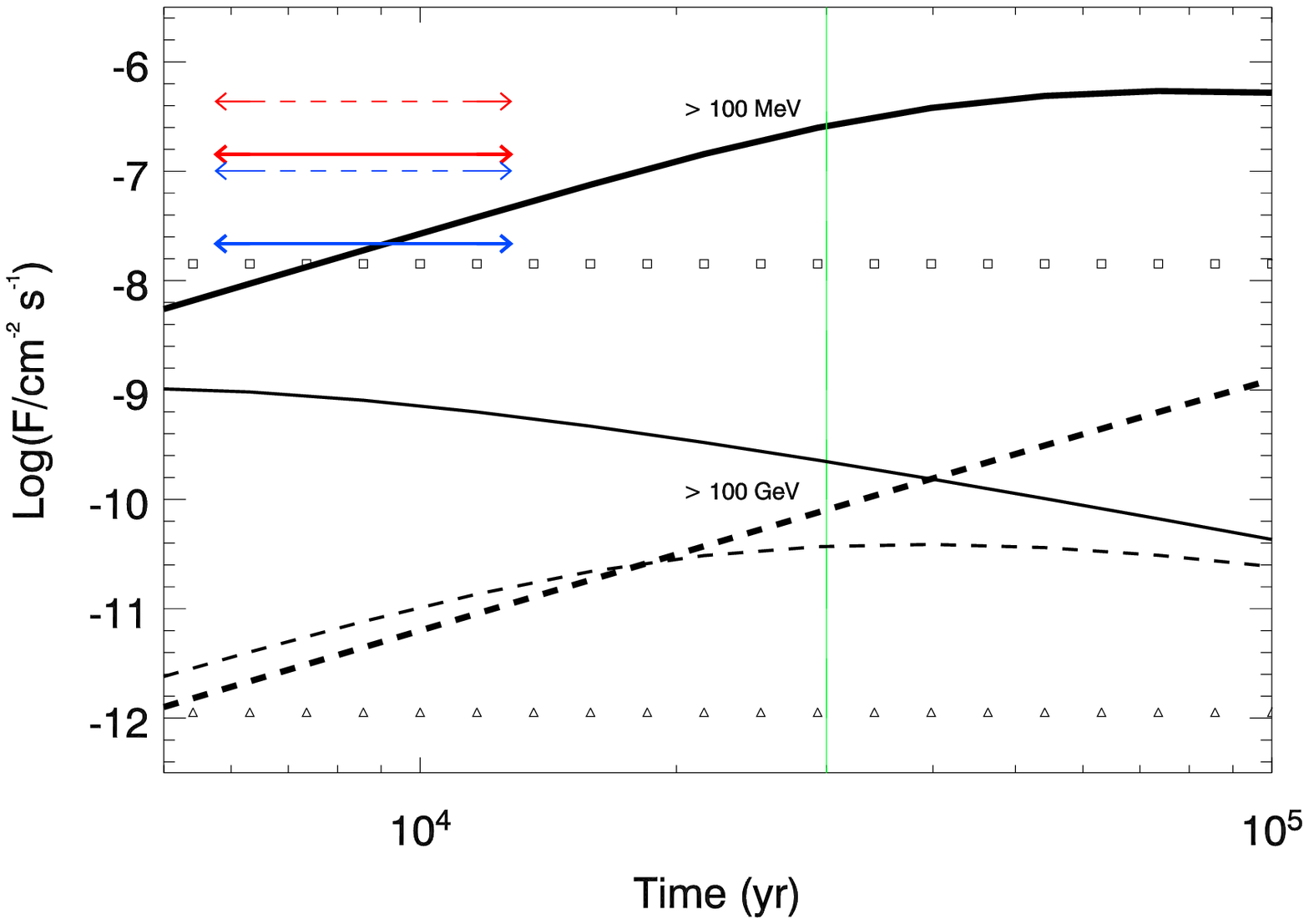}
\caption{Integrated photon flux as a function of time 
above 100 MeV and 100 GeV, solid (dashed) lines correspond to the case of the cloud located at 10 (30) pc. The horizontal lines represent the values of integrated fluxes in the case that the CR spectrum interacting with the cloud is the one found near Earth. The vertical line stands for the SNR age. EGRET and Fermi integral sensitivity, consistent in value and color coding with those in Fig. 2, are shown.}
\end{figure}

Fig. 2 also presents the results of our theoretical model focusing in the energy range of EGRET. There,  the CR spectrum interacting with a local-to-the-SNR cloud is obtained assuming an average distance of interaction of 3--4 pc. A few hundred M$_\odot$ located at this distance ($\sim$700 M$_\odot$ for the case of an impulsive, and $\sim$300 M$_\odot$ for a continuous case) produce an excellent match to the EGRET data, without generating a co-spatial MAGIC source. Concurrently with, e.g., Gaiser et al. (1999), we find that 
the lowest energy data points in the EGRET range are produced by bremsstrahlung of 
accelerated electrons, curves that for simplicity we do not show in Fig. 2. 

As spinoff of  the constraints provided by the observed phenomenology (e.g., the molecular environment and the position of the $\gamma$-ray sources) in the setting of this model, we find that $D_{10}$ should be low, of the order of 10$^{26}$ cm$^2$ s$^{-1}$. By varying the diffusion coefficient and studying its influence in our results, we obtain that if the separation between the giant cloud and the SNR is $>$10 pc, an slower diffusion would not allow sufficient high energy particles to reach the target material; thus, the MAGIC source would not be there. On the other hand, if the 
separation between the main cloud and the SNR is $<$10 pc, we would have detected an EGRET source at the position of the cloud, which is not the case. We then grasp the value of $D_{10}$ at 1.5 kpc from Earth, combining MAGIC and EGRET observations. Such values of $D_{10}$ are expected in dense regions of ISM such as the one we study (Ormes et al. 1988, Gabici \& Aharonian 2007).


\section{Predictions for Fermi}

The sensitivity of Fermi allows the observation of the previous scenario, but this observation also depends on the spatial resolution of the different clouds. Fig. 4 shows the predicted energy spectrum for the two cloud systems when Fermi is not able to resolve them independently. The final spectrum is the result of the addition of the two spectra in Fig. 3. The figure also depicts the best fit of the model to a power law. Table 1. shows the different spectral index for the different energy bands.
 
At high energies, we should see a morphological and a spectral change from the position of the cloud (i.e. the center of MAGIC J0616+225) towards the center of IC 443. At a morphological level, the lower the energy, the more coincident with the SNR the radiation will be detected. At a spectral level: sufficient statistics should show that the lower the $\gamma$-ray energy the harder the spectrum is. Fermi observations may also be sensitive enough to detect the same cloud that shines at higher energy, which ultimately will allow to determine its separation from the remnant, if the diffusion coefficient is assumed --as we showed--, or viceversa.

\begin{figure}
\centering
\includegraphics[width=0.8\columnwidth,trim=0 5 0 10]{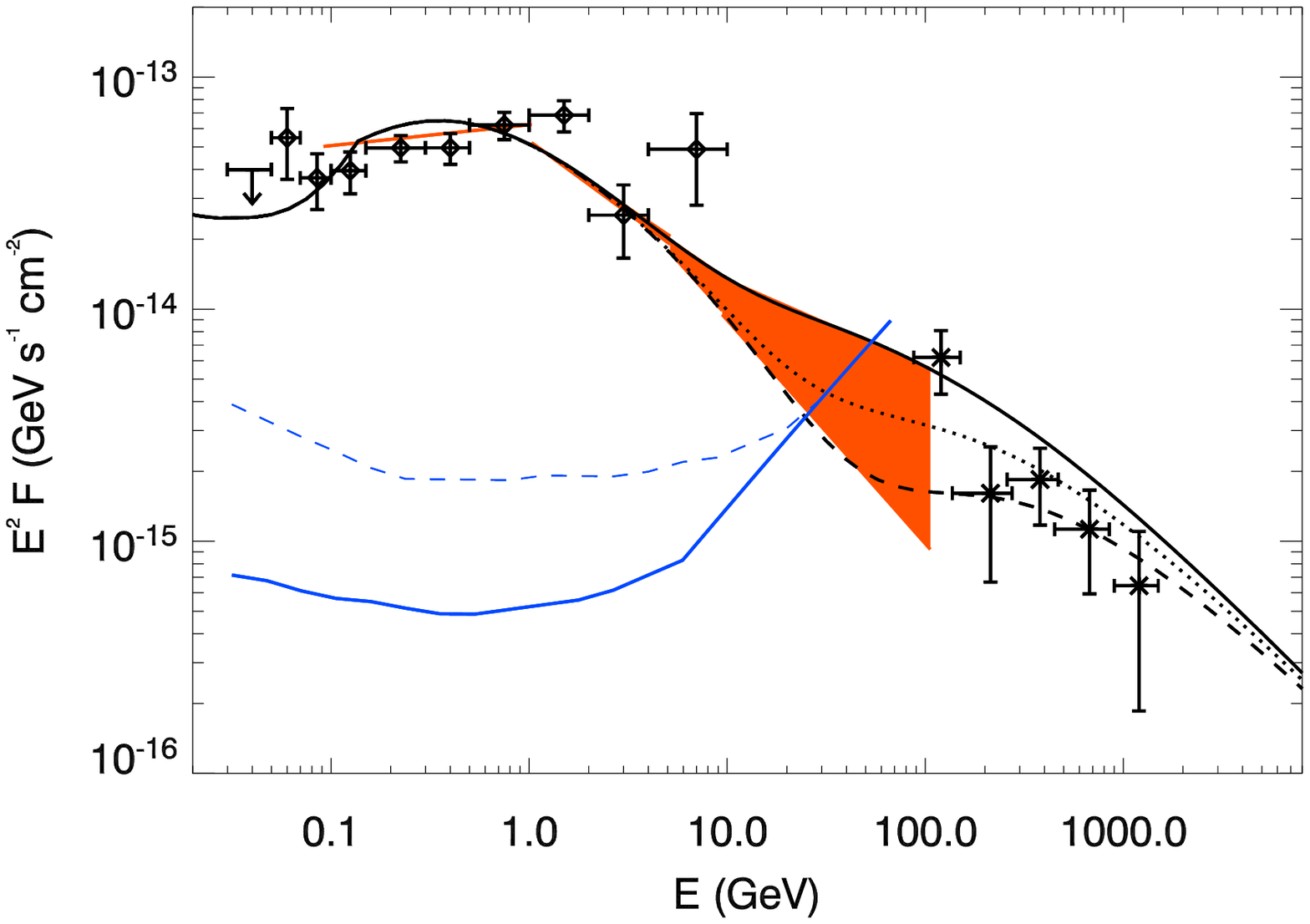}
\caption{MAGIC and EGRET measurement of the neighborhood of IC 443 (stars and squares, respectively) as compared with model predictions. 
The panel shows the results for an impulsive case where two curves have been added. This will be the scenario for those sources that Fermi can not resolve independently. 
At the MAGIC energy range, the curves show the predictions for a cloud of 8000 M$_\odot$ located at 20, 25, and 30 pc. At the EGRET energy range, the curve shows the prediction for a few hundred M$_\odot$ located at 3-4 pc.
The orange region show the best fit of the model to a power-law spectrum, and such way how the spectral index can depend on the different energy bands.
The Fermi sensitivity curves (in blue) (taken 
from http://www-glast.slac.stanford.edu/software/IS/glast\_latperformance.html)
show the  1-year sky-survey sensitivity for the Galactic North pole, again a position with low diffuse 
emission (solid), and for a typical position in the Inner Galaxy (dashed).}
\end{figure}

\begin{table}
\centering
  \caption{Mean values of the spectral indeces for the different energy ranges of the best fit to a power law of the model explaining the separation between MAGIC and EGRET sources.}
   \begin{tabular}{ccccc}
    \hline
     & $0.1$ $<$ E $<$ 1 & $1$ $<$ E $<$ 5 & $5$ $<$ E $<$ 10 & $10$ $<$ E $<$ 100 \\
    \hline
    Spectral index & -1.91 $\pm$ 0.01 &  -2.60 $\pm$ 0.03 & -2.77 $\pm$ 0.18 &  -2.66 $\pm$ 0.29  \\
    \end{tabular}               
    \end{table}

\section{Concluding remarks}

Here we have shown that MAGIC J0616+225 is consistent with the interpretation of CR interactions with a giant molecular cloud lying in front of the remnant, producing no counterpart at lower energies. We have also shown that the nearby EGRET source can be produced by the same accelerator, and that in this case, a co-spatial MAGIC source is not expected.  
In our model, the displacement between EGRET and MAGIC sources has a physical origin. It is generated by the different properties of the proton spectrum at different locations, in turn produced by the diffusion of CRs from the accelerator (IC 443) to the target. Specific predictions for future observations can be made as a result of this model as we have shown for the Fermi case.

\begin{theacknowledgments}

We acknowledge support by grants MEC-AYA 2006-00530 and CSIC-PIE 200750I029. The work of E. de Cea del Pozo has been made under the auspice of a FPI Fellowship, grant BES-2007-15131.  

\end{theacknowledgments}



\begin{thebibliography}{}

\bibitem{1.} Aharonian F. A., \& Atoyan A.M. 1996, A\&A 309, 91

\bibitem{2.} Albert J. et al. 2007, ApJ 664, L87
  
\bibitem{3.} Atoyan A. M., Aharonian F. A., \& V\"olk H. J. 1995, Phys.
Rev. D52, 3265  

\bibitem{4.} Asaoka I. \& Aschenbach B., 1994, A\&A 284, 573

\bibitem{5.} Bocchino F. \& Bykov A. M.
 2000, A\&A 362, L29

\bibitem{6.} Bocchino F. \& Bykov, A. M.
 2001, A\&A 376, 248

\bibitem{7.} Bocchino F. \& Bykov, A. M.
 2003, A\&A 400, 203

\bibitem{8.} Braun R. \& Strom R. J. 1986, A\&A, 164, 193

\bibitem{9.} Butt Y. M. et al. 2003, XXII Moriond Astro. Mtg, p. 323, astro-ph/0206132

\bibitem{10.} Bykov A. M., Bocchino F. \& Pavlov G. G. 
2005, ApJ 624, L41

\bibitem{11.} Bykov A. M. et al. 
2008, to appear in ApJ, arXiv:0801.1255

\bibitem{12.} Chevalier R. A. 1999,  ApJ  511, 798

\bibitem{13.} Claussen M. J. et al. 1997, ApJ 489, 143

\bibitem{14.} Cornett R. H., Chin G. \& Knapp G. R.,
 1977, A\&A 54, 889

\bibitem{15.} DeNoyer L.K. 1979, ApJ 232, L165

\bibitem{16.} Dickman R. L., Snell R. L., Ziurys L. M., \& Huang
Y.-L. 1992, ApJ 400, 203

\bibitem{17.} Domingo-Santamar\'{\i}a E. \& Torres D. F. 2005, A\&A 444, 403
  

\bibitem{18.} Fesen R. A. \& Kirshner R. P. 1980,  ApJ 242, 1023
  
\bibitem{19.}
 Funk S., Reimer O., Torres D. F. \& Hinton J. A. 2008, to appear in ApJ, 
  arXiv:0710.1584  
  
\bibitem{20.} Gabici  S. \& Aharonian F. A. 2007, ApJ 665, L131 
  
\bibitem{21.} Gaensler B. M. et al.  2006, ApJ, 648, 1037
  
\bibitem{22.} Gaisser T. K. 1990, Cosmic Rays and Particle Physics,
Cambridge University Press, Cambridge, UK
  
\bibitem{23.} Ginzburg V. L., Syrovatskii S. I. 1964, ``The Origin of
Cosmic Rays'', Pergamon Press, London
  
\bibitem{24.} Green D. A.  2004, BASI, 32, 335G
  
\bibitem{25.} Hartman R. C. et al. 1999, ApJS, 123, 79

\bibitem{26.} Hewitt J. W. et al.  2006, ApJ, 652, 1288
  
\bibitem{27.} Huang Y.-L., Dickman R. L., \& Snell R. L. 1986, ApJ 302, L63
  
\bibitem{28.} Kelner S. R., Aharonian F. A. \& Bugayov V. V. 
 2006, Phys. Rev. D74, 034018
  
\bibitem{29.} Keohane J. W. et al, 1997, ApJ 484, 350
    
\bibitem{30.} Lasker B. M. et al.  1990, AJ, 99, 2019
 
\bibitem{31.} Leahy D. A. 2004, AJ 127, 2277

\bibitem{32.} Lozinskaya T. A. 1981, Soviet Astron. Lett., 7, 17 

\bibitem{33.}  Olbert Ch. M. et al. 2001, ApJ 554, L205

\bibitem{34.} Ormes  J. F., Ozel M. E. \& Morris D. J. 1988, ApJ 334, 722

\bibitem{35.} 	Rho J., Jarrett T. H., Cutri R. M. \& Reach W. T.
2001,  ApJ 547, 885
  
\bibitem{36.} Rosado M., Arias L. \& Ambrocio-Cruz P. 2007, ApJ 133, 89
  
\bibitem{37.} Seta M., et al.  1998, ApJ, 505, 286

 \bibitem{38.} Torres D. F., Pessah M. E. \& Romero G. E. 2001, Astron. Nachr. 322, 223

\bibitem{39.} Torres D. F. et al. 2003, Phys. Rept., 382, 303.
 
\bibitem{40.} Torres D. F. 2004, ApJ 617, 966

\bibitem{41.} Troja E., Bocchino F. \& Reale F., 2006, ApJ, 649, 258

\end{thebibliography}
\end{document}